\documentclass{ijuc}

\usepackage{graphics}
\usepackage[cmex10]{amsmath}
\usepackage{amsthm,amssymb}
\usepackage{amsfonts}
\usepackage {url}
\usepackage{epstopdf}
\usepackage{float}
\usepackage{algorithm}
\usepackage{algorithmic}

\begin{document}
\title{End-to-End Delay Modeling for Mobile Ad Hoc Networks: A Quasi-Birth-and-Death Approach}

\author{Juntao Gao\inst{1}\email{gaojuntao223@gmail.com}
\and Yulong Shen\inst{2}\email{ylshen@mail.xidian.edu.cn}
\and Xiaohong Jiang\inst{1}\email{jiang@fun.ac.jp}
\and Osamu Takahashi\inst{1}
\and Norio Shiratori\inst{3}
}

\institute{School of Systems Information Science, Future University Hakodate, 041-8655 Japan
\and
School of Computer Science and Technology, Xidian University, 710071 China
\and GITS, Waseda University, Tokyo and RIEC, Tohoku University, Sendai-shi, 980-8579 Japan.
}

%

\maketitle

\begin{abstract}
Understanding the fundamental end-to-end delay performance in mobile ad hoc networks (MANETs) is of great importance for supporting Quality of Service (QoS) guaranteed applications in such networks.
While upper bounds and approximations for end-to-end delay in MANETs have been developed in literature, which usually introduce significant errors in delay analysis, the modeling of exact end-to-end delay in MANETs remains a technical challenge. This is partially due to the highly dynamical behaviors of MANETs, but also due to the lack of an efficient theoretical framework to capture such dynamics. 
This paper demonstrates the potential application of the powerful Quasi-Birth-and-Death (QBD) theory in tackling the challenging issue of exact end-to-end delay modeling in MANETs. We first apply the QBD theory to develop an efficient theoretical framework for capturing the complex dynamics in MANETs. We then show that with the help of this framework, closed form models can be derived for the analysis of exact end-to-end delay and also per node throughput capacity in  MANETs. Simulation and numerical results are further provided to illustrate the efficiency of these QBD theory-based models as well as our theoretical findings.
\end{abstract}


\section{Introduction}
Mobile ad hoc networks (MANETs) represent a class of important wireless ad hoc networks with mobile nodes. Since the flexible and distributed MANETs are robust and rapidly deployable/reconfigurable, they are highly appealing for a lot of critical applications \cite{Andrews_CM08,Goldsmith_CM11}, like deep space communication, disaster relief, battlefield communication, outdoor mining, device-to-device communication for traffic offloading in cellular networks, etc. 
To facilitate the applications of MANETs in providing Quality of Service (QoS) guaranteed services, understanding the fundamental delay performance of such networks is of great importance \cite{Hanzo_Survey07}.

Notice that the end-to-end delay, the time it takes a packet to reach its destination after it is generated at its source, serves as the most fundamental delay performance for a network.
However, the end-to-end delay modeling in MANETs remains a technical challenge. This is partially due to the highly dynamical behaviors of MANETs, like node mobility, interference, wireless channel/traffic contention, packet distributing, packet queueing process in a node and the complicated packet delivering process among mobile nodes, but also due to the lack of a theoretical framework to efficiently depict the complicated network state transitions under these network dynamics. 
By now, the available works on end-to-end delay analysis in MANETs mainly focus on deriving upper bounds or approximations for such delay. 

Based on the $M/G/1$ queueing model, some closed-form upper bounds on the expected end-to-end delay were derived for MANETs with two-hop relay routing \cite{Neely_IT05}.
For MANETs with multi-hop back-pressure routing, the Lyapunov drift model was adopted to derive an upper bound on the expected end-to-end delay \cite{Alresaini_INFOCOM12}. For MANETs with multi-hop linear routing, a network calculus approach was proposed to derive upper bounds on end-to-end delay distribution \cite{Ciucu_Allerton10, Ciucu_SIGMETRICS11}.
In addition to the delay upper bound results, approximations to end-to-end delay in MANETs have also been explored recently \cite{Hanbali_ASMTA08,Jindal_TMC09}. 
By adopting the polling model, an approximation to expected end-to-end delay was provided in \cite{Hanbali_ASMTA08} for a simple two-hop relay MANET consisting of only one source node, one relay node and one destination node. For more general MANETs with multiple source-destination pairs and multi-hop relay routing, approximations to corresponding end-to-end delay were developed in \cite{Jindal_TMC09} based on the  elementary probability theory.

It is notable that the results in \cite{Neely_IT05,Alresaini_INFOCOM12,Ciucu_Allerton10, Ciucu_SIGMETRICS11,Hanbali_ASMTA08,Jindal_TMC09} indicate that although above upper bound and approximation results are helpful for us to understand the general delay behaviors in MANETs, they usually introduce significant errors in end-to-end delay analysis. This is mainly due to the lack of an efficient theoretical framework to capture the complex network dynamics and thus the corresponding network state transitions in MANETs. 
This paper demonstrates the potential application of the powerful Quasi-Birth-and-Death (QBD) theory in capturing the network state transitions and thus in tackling the challenging exact end-to-end delay modeling issue in MANETs.

The main contributions of this paper are as follows:

\begin{itemize}

\item
We first demonstrate that the QBD theory actually enables a novel and powerful theoretical framework to be developed to efficiently capture the main network dynamics and thus the complex network state transitions in two-hop relay MANETs.

\item

With the help of the theoretical framework, we then show that we are able to analytically model the exact expected end-to-end delay and also exact per node throughput capacity in the concerned MANETs.

\item
Extensive simulation and numerical results are further provided to validate the efficiency of our QBD theory-based models for end-to-end delay and per node throughput capacity, and to illustrate how end-to-end delay and throughput capacity in MANETs are affected by some main network parameters.

\end{itemize}

The rest of this paper is organized as follows. Section~\ref{section:system_model} introduces the system models involved in this study. A QBD-based theoretical framework is developed in Section~\ref{section:delay_modeling} to capture network state transitions, based on which the exact expected end-to-end delay and per node throughput capacity are then derived in Section~\ref{section:e2e}.
Section~\ref{section:simulation_results} first provides simulation results to validate the efficiency of our theoretical framework and related delay and capacity results, and then explores the effects of network parameters on delay and capacity. Finally, we conclude this paper in Section~\ref{section:conclusion}.

\section{System Models} \label{section:system_model}
In this section, we introduce first the basic network models regarding node mobility, wireless channel, radio and traffic pattern in the considered MANET, and then Medium Access Control (MAC) protocol for transmission scheduling to resolve wireless channel contention and interference issues. Finally, the two-hop relay routing scheme that deals with packet delivery and traffic contention is discussed.

\subsection{Network Models}	\label{subsection:network-models}

\begin{figure}
\begin{center}
\scalebox{0.4}{\includegraphics{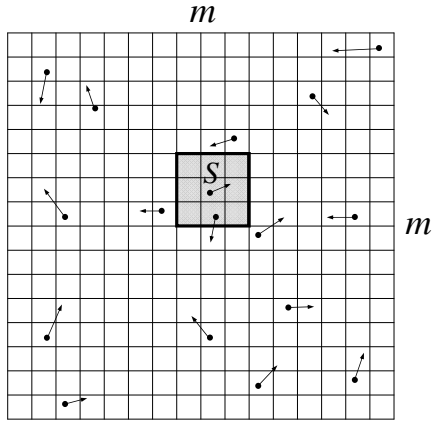}}
\end{center}
\caption{A snapshot of a cell partitioned MANET with $m=16$.}
\label{fig:network_partition}
\end{figure}

\textbf{Node mobility and channel models:} As shown in Fig.~\ref{fig:network_partition} that we consider a unit torus MANET partitioned evenly into $m\times m$ cells \cite{Sharma_TON07,Ciullo_TON11,Lipan_TMC12}. In the concerned MANET, there are $n$ nodes moving around according to the i.i.d. mobility model \cite{Grossglauser_TON02,Neely_IT05}. We consider the time slotted system, where each node randomly chooses one cell to move into at the beginning of every time slot and then stays in it for the whole time slot. We assume that a common bandwidth limited wireless channel is shared by all nodes for data transmissions. In each time slot, the data transmitted between any two nodes through the wireless channel is normalized to one packet.

\textbf{Radio model:}	Each node employs the same radio power to transmit data through the common wireless channel. To enable the transmission region of a node (say $S$ in Fig.~\ref{fig:network_partition}) to cover its own cell and also its $8$ neighbor cells (called coverage cells of the node hereafter), the corresponding radio range $r$ of the node should be set as $r=\sqrt{8}/m$. 
Based on the widely used protocol model \cite{Gupta_IT00,Kulkarni_IT04,Ciullo_TON11,Lipan_TMC12},
data transmission from a transmitting node (transmitter) $i$ to a receiving node (receiver) $j$ can be conducted only if the Euclidean distance $d_{ij}$ between them is less than $r$ (i.e., $d_{ij} \leq r$), while the data can be successfully received by receiver $j$ only if $d_{kj} \geq (1+\Delta)\cdot r$ holds for any other concurrent transmitter $k$, here $k \neq i,j$, and $\Delta \geq 0$ is a specified guard-factor for interference prevention.

\textbf{Traffic model:} Similar to previous work \cite{Ciullo_TON11}, we consider the permutation traffic pattern in which each node acts as the source of a traffic flow and at the same time the destination of another traffic flow. Thus, there are in total $n$ distinct traffic flows  in the MANET. Each source node exogenously generates packets for its destination according to an Bernoulli process with average rate $\lambda$ (packets/slot) \cite{Neely_IT05}.

\subsection{MAC Protocol}
\begin{figure}
\begin{center}
\scalebox{0.4}{\includegraphics{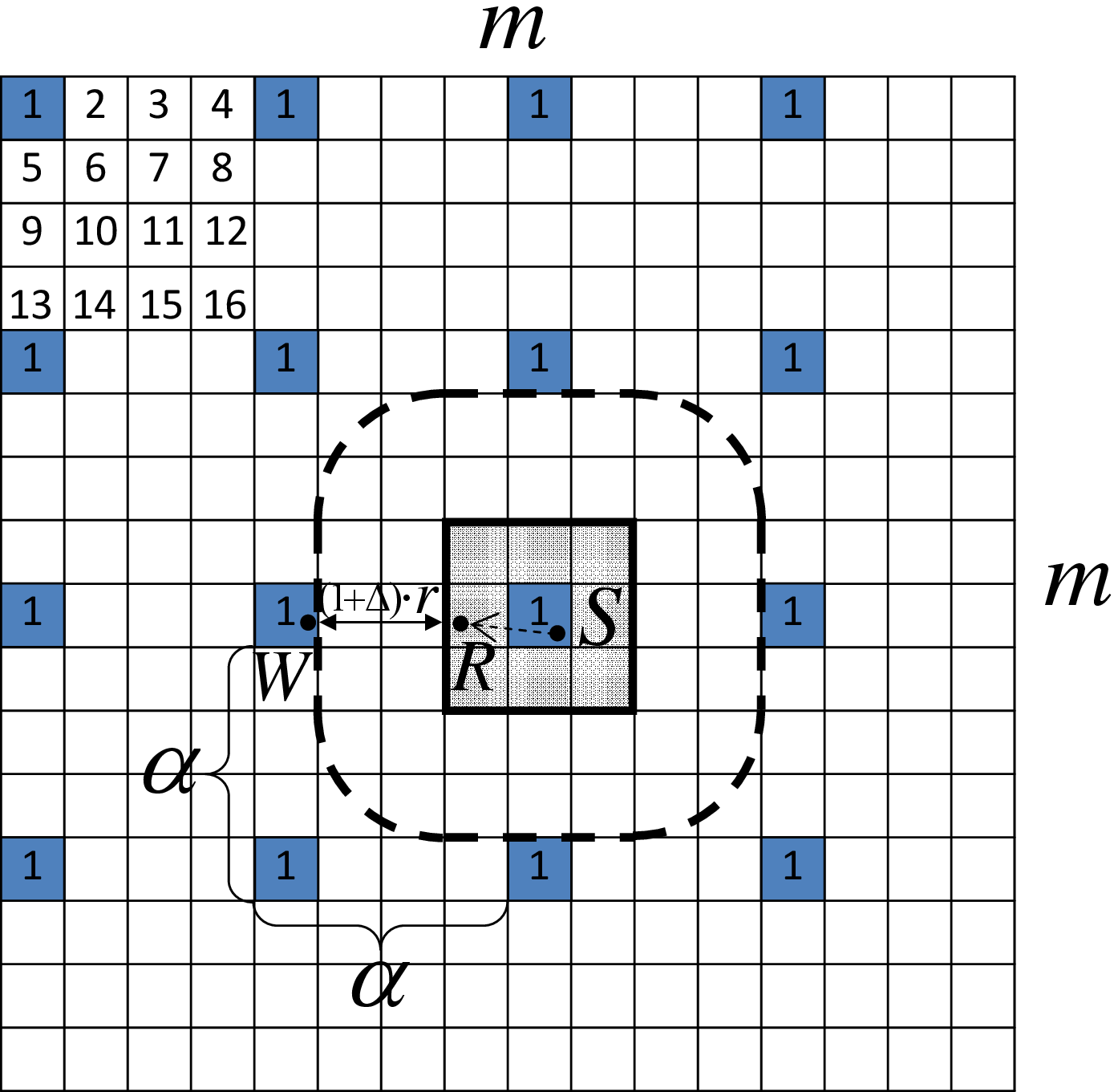}}
\end{center}
\caption{Illustration of equivalent-classes in a cell partitioned MANET. There are $16$ equivalent-classes in this MANET with $\alpha=4$. All shaded cells belong to equivalent-class $1$.}
\label{fig:network_EC}
\end{figure}

For a fair channel access, we consider here a commonly used MAC protocol for transmission scheduling, based on the idea of equivalent-class (EC) \cite{Kulkarni_IT04,Ciullo_TON11,Lipan_TMC12}. An EC is defined as a set of cells as illustrated in Fig.~\ref{fig:network_EC}, where any two cells in an EC are separated by a horizontal and vertical distance of some integer multiple of $\alpha$ cells ($1 \leq \alpha \leq m$). Thus, we have in total $\alpha^2$ ECs in the MANET. Under the EC based MAC protocol (MAC-EC), these ECs are scheduled to be active alternatively as time evolves. The cells in an active EC are called active cells and nodes (if any) in an active cell contend fairly to access the common wireless channel.

%
 
To enable as many number of concurrent transmissions to be scheduled as possible while avoiding interference among these transmissions, the parameter $\alpha$ should be set appropriately. As illustrated in Fig~\ref{fig:network_EC} that for the transmission between a transmitter $S$ and its possible receiver $R$ to be successful, the distance between $R$ and another possible closest concurrent transmitter $W$, i.e., $(\alpha-2)/m$, should satisfy the following condition according to the protocol model \cite{Gupta_IT00}:

\begin{align}
(\alpha-2)/m &\geq (1+\Delta)\cdot r,
\end{align}
Notice also that $r=\sqrt{8}/m$ and $\alpha \leq m$, the parameter $\alpha$ should be determined as 
\begin{align}
\alpha &= \min\{\lceil (1+\Delta)\sqrt{8}+2\rceil,m\},
\end{align}
where $\lceil x \rceil$ takes the least integer value greater than or equal to $x$. 


\subsection{Two-Hop Relay Routing} \label{subsection:2HR-B}

Once a node, say $S$ in Fig.~\ref{fig:network_EC}, succeeds in wireless channel contention and becomes a transmitter, it 
executes the popular two-hop relay (2HR) routing protocol defined in Algorithm~\ref{algorithm:2HR-B} for packet delivery \cite{Sadjadpour_TC07,Altman_TAC11}. With the 2HR routing, each exogenously generated packet at $S$ is first distributed out to relays through wireless broadcast \cite{Sadjadpour_TC07,Neely_AdHoc09}, and it is then delivered to its destination $D$ via these relays.

\floatname{algorithm}{Algorithm}

\begin{algorithm}[!ht]
\caption{2HR Routing Protocol}
\label{algorithm:2HR-B}
\begin{algorithmic}[1]
	\STATE Transmitter $S$ selects to conduct packet-broadcast with probability $q$, $0 < q < 1$, and to conduct packet-delivery with probability $1-q$;
	
	\IF{$S$ selects packet-broadcast}
		\STATE $S$ executes Procedure~\ref{procedure:packet-broadcast};
	\ELSE
		\STATE $S$ executes Procedure~\ref{procedure:packet-delivery};
	\ENDIF

\end{algorithmic}
\end{algorithm}

\floatname{algorithm}{Procedure}
\begin{algorithm}[!ht]
\renewcommand{\thealgorithm}{1.1}
\caption{packet-broadcast}
\label{procedure:packet-broadcast}
\begin{algorithmic}[1]
	\IF{$S$ has packets in its source-queue}
		\STATE $S$ distributes out the head-of-line (HoL) packet of source-queue through wireless broadcast to all nodes in its coverage cells;
		\STATE Any node, say $R$, in the coverage cells of $S$ reserves a copy of that packet;
		\IF{$R$ is not the destination $D$}
			\STATE $R$ inserts the HoL packet into the end of its relay-queue associated with $D$;
		\ELSE
			\IF{$R$ is currently requesting the HoL packet}
				\STATE $R$ keeps the HoL packet and increases $ACK(D)$ by $1$;
			\ELSE
				\STATE $R$ discards that packet;
			\ENDIF
		\ENDIF
		\STATE $S$ moves that HoL packet out of source-queue and inserts it into the end of its broadcast-queue;
		\STATE $S$ moves ahead the remaining packets in its source-queue;
	\ELSE
		\STATE $S$ remains idle;
	\ENDIF
\end{algorithmic}
\end{algorithm}

\floatname{algorithm}{Procedure}
\begin{algorithm}[!ht]
\renewcommand{\thealgorithm}{1.2}
\caption{packet-delivery}
\label{procedure:packet-delivery}
\begin{algorithmic}[1]
	\STATE $S$ randomly selects a node $U$ as its receiver from nodes in its coverage cells. Denote the source of $U$ as $V$;
	\STATE $S$ initiates a handshake with $U$ to acquire the packet number $ACK(U)+1$ and thus to know which packet $U$ is currently requesting;
	\STATE $S$ checks its corresponding relay-queue/broadcast-queue whether it bears a packet with $ID(V)=ACK(U)+1$;
	\IF{$S$ bears such packet}
		\STATE $S$ delivers that packet to $U$;
		\STATE $S$ clears all packets with $ID(V) \leq ACK(U)$ from its corresponding relay-queue/broadcast-queue;
		\STATE $S$ moves ahead the remaining packets in its corresponding relay-queue/broadcast-queue;
		\STATE $U$ increases $ACK(U)$ by $1$;
	\ENDIF

\end{algorithmic}
\end{algorithm}

To facilitate the operation of the 2HR routing protocol, each node, say $S$, is equipped with three types of First In First Out (FIFO) queues: one source-queue, one broadcast-queue and $n-2$ parallel relay-queues (no relay-queue is needed for node $S$ itself and its destination node $D$). 

\textbf{Source-queue:} Source-queue stores packets exogenously generated at $S$ and destined for $D$. These exogenous packets will be distributed out to relay nodes later in FIFS way.

\textbf{Broadcast-queue:} Broadcast-queue stores packets from source-queue that have already been distributed out by $S$ but have not been  acknowledged yet by $D$ the reception of them.

\textbf{Relay-queue:} Each node other than $S$ and $D$ is assigned with a relay-queue in $S$ to store redundant copies of packets distributed out by the source of that node.

To ensure the in-order packet reception at $D$,  similar to previous work \cite{Neely_IT05} that $S$ labels every exogenously generated packet with a unique identification number $ID(S)$, which increases by $1$ every time a packet is generated; destination $D$ also maintains an acknowledgment number $ACK(D)$ indicating that $D$ is currently requesting the packet with $ID(S)=ACK(D)+1$ (i.e, the packets with $ID(S) \leq ACK(D)$ have already been received by $D$).

\section{QBD-Based Theoretical Framework} \label{section:delay_modeling}
In this section, we first present some preliminaries, and then develop a novel theoretical framework based on the QBD theory to capture the complex network state transitions in the concerned MANETs. 

\subsection{Preliminaries}
We focus on one specific traffic flow from source $S$ to destination $D$ in our analysis. 
Notice that once a packet is generated at $S$, it first experiences a queueing process in the source-queue of $S$ before being distributed out (served), and it then experiences a network delivery process after being distributed out into the network by $S$ and before being successfully received by $D$. 
Since $D$ requests packets in order according to $ACK(D)$, all packets distributed out by $S$ will be also delivered (served) in order. Thus, we can treat the network delivery process as a queueing process of one virtual network-queue. Notice also that the departure process of source-queue is just the arrival process of network-queue.

To fully depict the two queueing processes in both source-queue and network-queue, we define following probabilities for a time slot.

\begin{itemize}

\item $p_{b}:$ probability that $S$ becomes transmitter and also selects to do packet-broadcast.

\item $p_{c}(j):$ probability that $j$ copies of a packet exist in the network (including the one in $S$) after the packet is distributed out by $S$ in the current time slot, $1 \leq j \leq n-1$.

\item $p_{r}(j):$ probability that $D$ receives the packet it is currently requesting given that $j$ copies of the packet exist in the network, $1 \leq j \leq n-1$.

\item	$p_{0}(j):$ probability that $j$ copies of a packet  exist in the network after $S$ becomes transmitter and selects to do packet-broadcast for this packet, given that $D$ is out of the coverage cells of $S$ and network-queue is empty, $1 \leq j \leq n-1$. 

\item	$p_{0}(0):$ $p_{0}(0)=1-\sum_{j=1}^{n-1}p_{0}(j)$.

\item	$p_{b}^{+}(j):$ probability that $S$ becomes transmitter, selects to do packet-broadcast and also successfully conducts packet-broadcast for one packet; at the same time, $D$ receives the packet it is requesting given that $j$ copies of that packet exist in the network, $1 \leq j \leq n-1$.

\item	$p_{b}^{-}(j):$	probability that $S$ becomes transmitter,  selects to do packet-broadcast and also successfully conducts packet-broadcast for one packet; at the same time, $D$ does not receive the packet it is requesting given that $j$ copies of that packet exist in the network, $1 \leq j \leq n-1$.

\item	$p_{f}^{+}(j):$ probability that $S$ does not successfully conduct packet-broadcast for any packet; at the same time, $D$ receives the packet it is requesting given that $j$ copies of that packet exist in the network, $1 \leq j \leq n-1$.

\item	$p_{f}^{-}(j):$ probability that $S$ does not successfully conduct packet-broadcast for any packet; at the same time, $D$ does not receive the packet it is requesting  given that $j$ copies of that packet exist in the network, $1 \leq j \leq n-1$.
\end{itemize}

The following lemma reveals a nice property about the source-queue and network-queue, which will help us to evaluate the above probabilities in Lemma~2. 

\emph{Lemma 1}:
For the considered MANET with MAC-EC protocol for transmission scheduling and 2HR-B protocol for packet delivery, the arrival process of network-queue is a Bernoulli process with probability $\lambda$ and it is independent of the state of source-queue.

\begin{proof}
We know from Section~\ref{subsection:network-models} that the arrival process of source-queue in $S$ is a Bernoulli process with probability $\lambda$. The service process of source-queue is actually also a Bernoulli process, because in every time slot  $S$ gets a chance with constant probability $p_{b}$ to do packet-broadcast to distribute out a packet in source-queue  (or equivalently, the source-queue is served with probability $p_{b}$ in every time slot). Thus, the source-queue in $S$ follows a Bernoulli/Bernoulli queue, and in equilibrium the packet departure process of source-queue is also a Bernoulli process with probability $\lambda$, which is independent of the state of source-queue (i.e., the number of packets in source-queue) \cite{Daduna_Book01}. Because the arrival process of network-queue is just the departure process of source-queue, this finishes the proof of this Lemma.
\end{proof}

\emph{Lemma 2}:

\begin{align}
p_b					&= \frac{qm^2}{\alpha^2n}\bigg\{ 1-\bigg(\frac{m^2-1}{m^2}\bigg)^n\bigg\}  \label{Pb}\\
p_{c}(j)		&= 	\!\frac{n\binom{n-2}{j-1}(m^2\!-\!9)^{n-1-j}}{m^{2n}\!-\!(m^2-1)^n}\Big\{\!(m^2-9)f(j)+\!f(j\!+\!1)\Big\}		\label{Pc(j)}	\\
p_{r}(j)		&=	\!\frac{j(1\!-\!q)m^2}{\alpha^2n(n\!-\!1)} \bigg\{\!	1\!-\!\bigg(\frac{m^2\!-\!1}{m^2}\bigg)^n\!-\!\!\frac{n}{m^2}\bigg(\frac{m^2-9}{m^2}\bigg)^{n\!-\!1}\bigg\}		\label{exp:pr(j)}  \\
p_{0}(j)	&=	\frac{\lambda \cdot q \cdot \binom{n-2}{j-1}(m^2-9)^{n-j}}{\alpha^2 m^{2n-2}p_{b}}f(j)\\
p_{0}(0)	&=	1- \frac{\lambda \cdot q \cdot (m^2-9)}{\alpha^2(n-1)p_{b}}\bigg\{	1-\bigg(\frac{m^2-1}{m^2}\bigg)^{n-1}\bigg\} \\
p_{b}^{+}(j)	&=(j-1)\frac{\lambda(q-q^2)(m^4-m^2\alpha^2)}{\alpha^4 n(n-1)(n-2) p_{b}}	\nonumber \\
							&		\quad \cdot \bigg\{1-2\bigg(\frac{m^2-1}{m^2}\bigg)^n + \bigg(\frac{m^2-2}{m^2}\bigg)^n	\nonumber \\
							&		\quad		-\frac{n}{m^2}\bigg(\frac{m^2-9}{m^2}\bigg)^{n-1}\!+\!\frac{n}{m^2}\bigg(\frac{m^2\!-\!10}{m^2}\bigg)^{n\!-\!1}\bigg\} \label{exp:pb+(j)} \\
p_{b}^{-}(j)	&=	\lambda-p_{b}^{+}(j)	\\
p_{f}^{+}(j)	&=	p_{r}(j)-p_{b}^{+}(j) \\
p_{f}^{-}(j)	&= 	1 - p_{b}^{+}(j) - p_{b}^{-}(j) - p_{f}^{+}(j)	\label{exp:pf-(j)}
\end{align}
where 
\begin{align}
f(x)=\frac{9^x-8^x}{x} 
\end{align}

\begin{proof}
The proof of Lemma~2 is given in Appendix~A.
\end{proof}

\emph{Remark}:
The complex network dynamics of node mobility, interference, wireless channel and traffic contention are incorporated into the calculation of the above probabilities as shown in Appendix~A. The network dynamics of packet distributing, packet queueing and delivering processes will be captured in the following QBD modeling process.

\subsection{QBD Modeling}

We use $L(t) \geq 0$ to denote the number of local packets distributed out from $S$ but not received yet by $D$ until time slot $t$, and use $J(t)$ to denote the number of copies of the packet $D$ is currently requesting at time slot $t$ in the network, $0 \leq J(t) \leq n-1$.
As time $t$ evolves, the queueing process of network-queue follows a two-dimensional QBD process \cite{Alfa_Book10,Latouche_Book99} 
\begin{align}	\label{markov-process}
\{(L(t), J(t)), t = 0, 1, 2, \cdots \}, 
\end{align}
on state space
\begin{align}	\label{markov-state-space}
\big\{ \{(0,0)\} \cup \{(l,j)\}; l \geq 1, 1 \leq j \leq n-1 \big\}
\end{align}
where  $(0,0)$ corresponds to the empty network-queue state. $L(t)$ increases by $1$ if $S$ distributes out a packet from its source-queue while $D$ does not receive the packet it is requesting at slot $t$, $L(t)$ decreases by $1$ if $S$ does not distribute out a packet from its source-queue while $D$ receives the packet it is requesting at slot $t$, and $L(t)$ keeps unchanged, otherwise. 

All states in (\ref{markov-state-space}) can be divided into the following subsets 
\begin{align}
N(0)	&=	\{(0,0)\}	\\
N(l)	&=	\big\{ \{(l,	j)\}, 1 \leq j \leq n-1 \big\}, l \geq 1
\end{align}
where subset $N(0)$ is called level $0$ and subset $N(l)$ is called level $l$. It is notable that when network-queue is in some state of level $l$ ($l \geq 1$) at a time slot, the next state of one-step state transitions could only be some state in the same level $l$ or in its adjacent levels $l-1$ and $l+1$. 

\begin{figure*}[!t]
\begin{center}
\scalebox{0.5}{\includegraphics{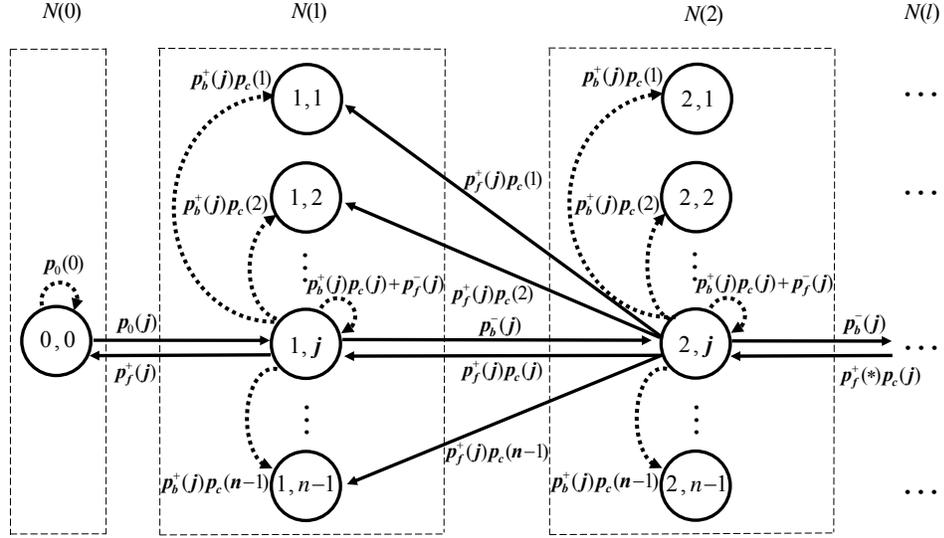}}
\end{center}
\caption{State transition diagram for the QBD process of network-queue.}
\label{fig:markov_process}
\end{figure*}

Based on the queueing process of network-queue and the definitions of probabilities in Lemma~2, the underlying QBD process of the network-queue has state transition diagram shown in Fig.~\ref{fig:markov_process}. In Fig.~\ref{fig:markov_process}, $p_{f}^{+}(\ast)p_{c}(j)$ denotes the probability of the transition from some state $(3, \ast)$ in level $3$ to the state $(2, j)$ in level $2$, where the asterisk `$\ast$' means some eligible copy number in $\{1, 2, \cdots, n-1\}$.

To facilitate our discussion, we classify the state transitions in Fig.~\ref{fig:markov_process} as \textbf{intra-level transition} (denoted by dotted arrows in Fig.~\ref{fig:markov_process}) and \textbf{inter-level transition} (denoted by solid arrows in Fig.~\ref{fig:markov_process}).

\textbf{Intra-level Transition:} There are two cases regarding the intra-level transitions, namely, the state transition inside level $0$ and the state transitions inside level $l$ ($l \geq 1$). For level $0$, it has only one state $(0,0)$, which could only transit to itself. For levels $l \geq 1$, they all follow the same intra-level transitions, i.e., a state $(l, j)$ in level $l$ could transit to any state (including itself) in the same level, $1 \leq j \leq n-1$.

\textbf{Inter-level Transition:} There are also two cases regarding the inter-level transitions, namely, transitions between level $0$ and level $1$, and transitions between level $l$ and level $l+1$ ($l \geq 1$). The inter-level transitions between level $0$ and level $1$ are simply bi-transitions between state $(0,0)$ and any state $(1, j)$, $1 \leq j \leq n-1$. For adjacent levels $l$ and $l+1$ ($l \geq 1$), they all follow the same inter-level transitions, i.e., a state $(l, j)$ in level $l$ could only transit to the corresponding state $(l+1, j)$ in level $l+1$, while a state $(l+1, j)$ in level $l+1$ could transit to any state in level $l$, $1 \leq j \leq n-1$. 

\section{Delay and Throughput Capacity} \label{section:e2e}
With the help of the QBD-based theoretical framework, we derive the expected end-to-end delay and also per node throughput capacity for the concerned MANETs.

\emph{Definition 1}:
End-to-end delay $T_{e}$ of a packet is the time elapsed between the time slot the packet is generated at its source and the time slot it is delivered to its destination.

\emph{Definition 2}:
Per node throughput capacity $\mu$ is defined as the maximum packet arrival rate $\lambda$ every node in the concerned MANET can stably support.

Before presenting our main result on the expected end-to-end delay, we first derive the per node throughput capacity, with which the input rate the MANET can stably support and the corresponding end-to-end delay can then be determined.

\emph{Theorem 1}:
For the considered MANET, its per node throughput capacity $\mu$ is given by
\begin{align}
\mu	&= \min \Bigg\{p_b, \frac{1}{\sum_{j=1}^{n-1} \frac{p_{c}(j)}{p_{r}(j)}}\Bigg\} \label{exp:throughput}
\end{align}

\begin{proof}
In equilibrium, the service rate $\mu_s$ of source-queue is 
\begin{align}
\mu_s	&= p_b
\end{align}
and the service rate $\mu_d$ of network-queue, i.e., the rate $D$ receives its requesting packets, is
\begin{align}
\mu_d	&=	\frac{1}{\sum_{j=1}^{n-1} \frac{p_{c}(j)}{p_{r}(j)}}.
\end{align}
To ensure network stability, packet generation rate $\lambda$	at $S$ should satisfy
\begin{align}
\lambda	&< \min \{\mu_s, \mu_d\}
\end{align}
Thus, the per node throughput capacity $\mu$ is determined as
\begin{align}
\mu	&= \min \{\mu_s, \mu_d\}
\end{align}
\end{proof}

Based on above per node throughput capacity result and the QBD-based theoretical framework, we now establish the following theorem on the expected end-to-end delay of the concerned MANET.

\emph{Theorem 2}:
For the concerned MANET, where each source node exogenously generates packets according to a Bernoulli process with probability $\lambda$ ($\lambda<\mu$), the expected end-to-end delay $\mathbb{E}(T_{e})$ of a packet is determined as 
\begin{align}	
\mathbb{E}(T_{e})	&=	\frac{\overline{L}_1+\overline{L}_2}{\lambda},	\label{exp:e2edelay}
\end{align}
where 
\begin{align}
\overline{L}_1	&=	\frac{\lambda-\lambda^2}{p_b-\lambda}	\\
\overline{L}_2	&=	\frac{\mathbf{y_1}(\mathbf{I}-\mathbf{R})^{-2}\mathbf{1}}{\phi}
\end{align}
\begin{align}	
\mathbf{R}					&=	\mathbf{A_0}(\mathbf{I}-\mathbf{A_1}-\mathbf{A_0} \mathbf{1}\mathbf{v_0})^{-1} \\
[y_0, \mathbf{y_1}]	&=	[y_0, \mathbf{y_1}]	\left[
																									\begin{array}{cc}
																									\mathbf{B_1}	&	\mathbf{B_0}	\\
																									\mathbf{B_2}	&	\mathbf{A_1}+\mathbf{R}\mathbf{A_2}
																									\end{array}
																						\right]	\label{exp:y1}	\\
\phi	&=	y_0+\mathbf{y_1}(\mathbf{I}-\mathbf{R})^{-1}\mathbf{1}	\label{exp:upsilon} \\
	\mathbf{v_0}	&=	\left[ \begin{array}{cccccc}
												\!\!p_{c}(1)	\!&\!	p_{c}(2)	\!&\! \cdots  \!&\! p_{c}(j) \!&\!	\cdots \!&\!	p_{c}(n\!-\!1)
											\end{array}
										\right]	\\
	\bf{A_0}	&= \mathrm{diag} \left(p_{b}^{-}(1), p_{b}^{-}(2), \cdots, p_{b}^{-}(j), \cdots, p_{b}^{-}(n\!-\!1)\right) \label{matrix-A0} \\
	\mathbf{A_1}	&=	\!\mathrm{diag} \big(p_{f}^{-}(1), p_{f}^{-}(2), \cdots, p_{f}^{-}(j), \cdots, p_{f}^{-}(n\!-\!1)\!\big)+ \nonumber \\
								&	 \quad \left[ \begin{array}{cccccc}
												\!\!p_{b}^{+}(1)	\!&\!	p_{b}^{+}(2)	\!&\! \cdots  \!&\! p_{b}^{+}(j) \!&\!	\cdots \!&\!	p_{b}^{+}(n\!-\!1)
														\end{array}
										\!\!\right]^T	\!\mathbf{v_0}	\label{matrix-A1}	\\
	\bf{A_2}	&= \mathbf{B_2 v_0}	\label{matrix-A2} \\
	\bf{B_0}	&=\left[ \begin{array}{cccccc}
												\!\!p_{0}(1)	\!&\!	p_{0}(2)	\!&\! \cdots  \!&\! p_{0}(j) \!&\!	\cdots \!&\!	p_{0}(n\!-\!1)
											\end{array}
							\!\right]	\label{matrix-B0} \\
	\bf{B_1}	&=\left[ p_{0}(0)\right]	\label{matrix-B1} \\
	\bf{B_2}	&=\left[ \begin{array}{cccccc}
												\!\!p_{f}^{+}(1)	\!&\!	p_{f}^{+}(2)	\!&\! \cdots  \!&\! p_{f}^{+}(j) \!&\!	\cdots \!&\!	p_{f}^{+}(n\!-\!1)
											\end{array}
							\!\right]^T	\label{matrix-B2}
\end{align}
here $\bf{I}$ denotes an identity matrix of size $(n-1)\times(n-1)$, $\mathbf{1}$ denotes a column vector of size $(n-1)\times 1$ with all elements being $1$, $y_0$ is a scalar value, and $\mathbf{y_1}$ is a row vector of size $1 \times (n-1)$.

\begin{proof}
From Lemma~1 we know that we can analyze queueing processes of source-queue and network-queue separately.

First, for the source-queue at $S$, since it follows a Bernoulli/Bernoulli queue, we know from \cite{Neely_IT05,Alfa_Book10} that the expected number of packets  $\overline{L}_1$  in the queue is determined as 
\begin{align}
\overline{L}_1	&= \frac{\lambda-\lambda^2}{p_b-\lambda}	
\end{align}

Then, for the network-queue, its queueing process follows a QBD process shown in Fig.~\ref{fig:markov_process}. The corresponding state transition matrix $\bf{Q}$ of the transition diagram in Fig.~\ref{fig:markov_process} is given by
\begin{align}	\label{state-matrix}
\bf{Q}=& \left[	\begin{array}{ccccc}
									\mathbf{B_1} &	\mathbf{B_0}	&	\mathbf{0}	&	\mathbf{0}	&	\cdots \\ 
									\bf{B_2} &	\bf{A_1}	&	\bf{A_0}	&	\bf{0}	&	\cdots \\
									\bf{0} 	&	\bf{A_2}	&	\bf{A_1}	&	\bf{A_0}	&	\cdots \\
									\bf{0} 	&	\bf{0}	&	\bf{A_2}	&	\bf{A_1}	&	\cdots \\
									\vdots 	&	\vdots	&	\vdots	&	\vdots	&	\ddots \\
								\end{array}
				\right]
\end{align}
where $\mathbf{B_{1}}$ defined in (\ref{matrix-B1}) represents the state transition from $(0,0)$ to $(0,0)$; $\mathbf{B_{0}}$ defined in (\ref{matrix-B0}) represents the state transitions from $(0,0)$ to $(1,j)$, $1 \leq j \leq n-1$; $\mathbf{B_{2}}$ defined in (\ref{matrix-B2}) represents the state transitions from $(1,j)$ to $(0,0)$, $1 \leq j \leq n-1$; $\mathbf{A_{1}}$ defined in (\ref{matrix-A1}) represents the state transitions from $(l,j)$ to $(l,i)$, $l \geq 1$, $1 \leq j,i \leq n-1$; $\mathbf{A_{0}}$ defined in (\ref{matrix-A0}) represents the state transitions from $(l,j)$ to the corresponding $(l+1,j)$, $l \geq 1$, $1 \leq j \leq n-1$; $\mathbf{A_{2}}$ defined in (\ref{matrix-A2}) represents the state transitions from $(l,j)$ to $(l-1,i)$, $l \geq 2$, $1 \leq j,i \leq n-1$.

Based on the QBD process theory \cite{Alfa_Book10,Latouche_Book99}, the queueing process of the network-queue can be analyzed through two related matrices $\mathbf{R}$ and $\mathbf{G}$ determined as:
\begin{align}
\mathbf{R}	&=	\mathbf{A_0}(\mathbf{I}-\mathbf{A_1}-\mathbf{A_0}\mathbf{G})^{-1} \\
\mathbf{G}	&=  \mathbf{A_2} + \mathbf{A_1}\mathbf{G}+ \mathbf{A_0}\mathbf{G}^2
\end{align}
where $\mathbf{R}=(r_{ij})_{(n-1)\times(n-1)}$, the entry $r_{ij}$ ($1 \leq i,j \leq n-1$) of matrix $\mathbf{R}$ is the expected number that the QBD of network-queue visits state ($l+1$, $j$) before it returns to states in $N(0) \cup \cdots \cup N(l)$, given that the QBD starts in state ($l$, $i$), and $\mathbf{G}=(g_{ij})_{(n-1)\times(n-1)}$, the entry $g_{ij}$ ($1 \leq i,j \leq n-1$) of matrix $\mathbf{G}$ is the probability that the QBD starts from state ($l$, $i$) and visits state ($l-1$, $j$) in a finite time.

Due to the special structure of $\mathbf{A_2}$, which is the product of a column vector $\mathbf{B_2}$ by a row vector $\mathbf{v_0}$, matrix $\mathbf{G}$ can be calculated as 
\begin{align}
\mathbf{G}	&=	\mathbf{1}\mathbf{v_0}
\end{align}

Based on the results in \cite{Alfa_Book10}, the expected number of packets  $\overline{L}_2$ of network-queue is given by 
\begin{align}
\overline{L}_2	&=	\frac{\mathbf{y_1}(\mathbf{I}-\mathbf{R})^{-2}\mathbf{1}}{\phi},
\end{align}
where $\mathbf{y_1}$ and $\phi$ are determined by (\ref{exp:y1}) and (\ref{exp:upsilon}), respectively.

Finally, by applying Little's Theorem \cite{Bertsekas_Book92},	(\ref{exp:e2edelay}) follows. This finishes the proof of Theorem~2.
\end{proof}

\section{Numerical Results}\label{section:simulation_results}
To validate the QBD-based theoretical results on expected end-to-end delay and per node throughput capacity, a customized C++ simulator has been developed to simulate packet generating, distributing and delivering processes in the considered MANET\footnote{The program of our simulator is now available online at \cite{2HR-B}. Similar to \cite{ns2}, the guard-factor is set as $\Delta=1$.}. In the simulator, not only the i.i.d. node mobility model but also the typical random walk \cite{Gamal_IT06} and random waypoint \cite{Zhou_Infocom10} mobility models have been implemented.
\begin{itemize}
\item{\textbf{Random Walk Model:}}
At the beginning of each time slot, each node first independently selects a cell with equal probability $1/9$ among its current cell and its $8$ neighboring cells; it then moves into that cell and stays in it until the end of that time slot.

\item{\textbf{Random Waypoint Model:}}
At the beginning of each time slot, each node first independently generates a two-element vector $[x, y]$, where both elements $x$ and $y$ are uniformly drawn from $[1/m , 3/m]$; it then moves along the horizontal and vertical direction of distance $x$ and $y$, respectively.

\end{itemize}

\subsection{End-to-End Delay Validation}

\begin{figure}[!t]
\begin{center}
\scalebox{0.8}{\includegraphics{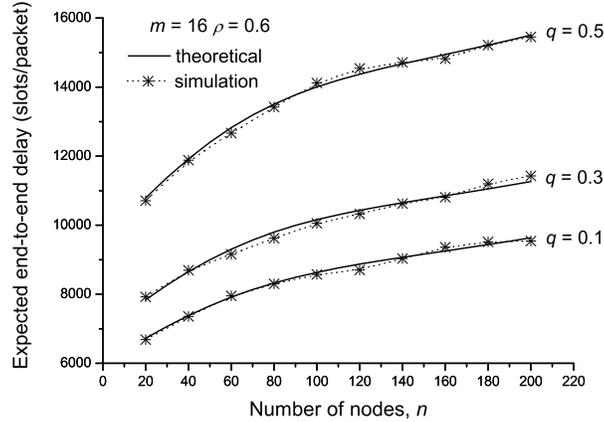}}
\end{center}
\caption{Expected packet end-to-end delay vs. number of nodes $n$ in MANET.}
\label{fig:e2edelay-vs-n}
\end{figure}

\begin{figure}[!t]
\begin{center}
\scalebox{0.8}{\includegraphics{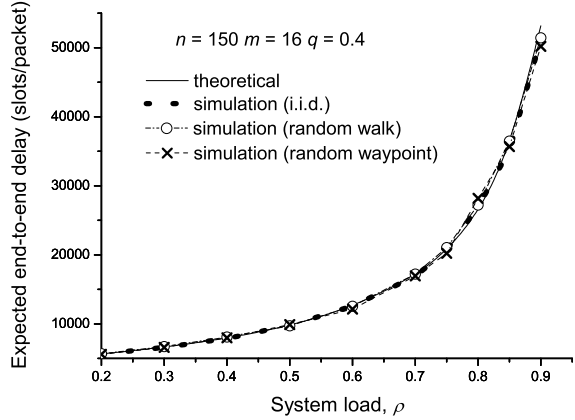}}
\end{center}
\caption{Expected packet end-to-end delay vs. system load $\rho$ in MANET.}
\label{fig:e2edelay-vs-rho}
\end{figure}

For networks of different size $n$, Fig.~\ref{fig:e2edelay-vs-n} shows both theoretical and simulation results on packet end-to-end delay under the settings of $m=16$,  system load $\rho=0.6$ $(\rho=\lambda/\mu)$ and packet-broadcast probability  $q = \{0.1, 0.3, 0.5\}$. Unless otherwise mentioned, simulation results are reported with small $95\%$ confidence intervals. The results in Fig.~\ref{fig:e2edelay-vs-n} show clearly that in a wide range of network scenarios considered here, theoretical results match very nicely with simulated ones, indicating that our QBD-based theoretical modeling is really efficient in capturing the expected packet end-to-end delay behavior of concerned MANETs. From Fig.~\ref{fig:e2edelay-vs-n} we can also see that as network size $n$ increases, packet end-to-end delay increases as well. This is because that in the concerned MANET with fixed unit area and fixed setting of $m=16$, as $n$ increases the contention for wireless channel access becomes more intensive, resulting in a lower packet delivery opportunity and thus a longer packet end-to-end delay.

For the setting of $n = 150, m = 16$ and $q = 0.4$, Fig.~\ref{fig:e2edelay-vs-rho} shows both the theoretical and simulation results on packet end-to-end delay when system load $\rho$ changes from $\rho=0.2$ to $\rho=0.9$. In addition to the i.i.d. mobility model considered in this paper, the corresponding simulation results for the random walk and random waypoint mobility models have also been included in Fig.~\ref{fig:e2edelay-vs-rho} for comparison. Again, we can see from Fig.~\ref{fig:e2edelay-vs-rho} that our theoretical delay model is very efficient. It is interesting to see from Fig.~\ref{fig:e2edelay-vs-rho} that although our theoretical framework is developed under the i.i.d. mobility model, it can also nicely capture the general packet end-to-end delay behavior under more realistic random walk and random waypoint mobility models.

\subsection{Throughput Capacity Validation}

\begin{figure}[!t]
\begin{center}
\scalebox{0.9}{\includegraphics{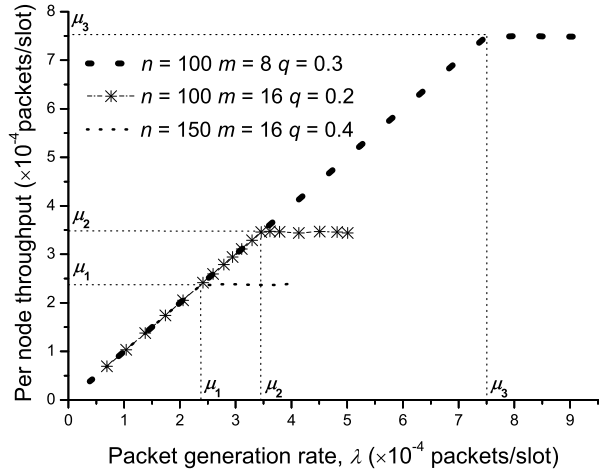}}
\end{center}
\caption{Per node throughput VS. packet generation rate $\lambda$ in MANET.}
\label{fig:throughput-vs-rho}
\end{figure}

Another observation of Fig.~\ref{fig:e2edelay-vs-rho} is that the packet end-to-end delay increases sharply as system load $\rho$ approaches $1.0$ (i.e., as packet generation rate $\lambda$ approaches per node throughput capacity $\mu$), which serves as an intuitive verification of our theoretical per node throughput capacity result.  To further validate our theoretical model on throughput capacity, Fig.~\ref{fig:throughput-vs-rho} provides the simulation results on the achievable per node throughput, i.e., the average rate of packet delivery to destination, when packet generation rate $\lambda$ increases gradually, where the results of three network scenarios with different throughput capacity $\{n = 150, m = 16, q = 0.4, \mu_1 = 2.37 \times 10^{-4}\}$, $\{n = 100, m = 16, q = 0.2, \mu_2 = 3.46 \times 10^{-4}\}$ and 
$\{n = 100, m = 8, q = 0.3, \mu_3 = 7.52 \times 10^{-4}\}$ are presented. We can see from Fig.~\ref{fig:throughput-vs-rho} that for each network scenario there, the corresponding per node throughput first increases monotonously as $\lambda$ increases before $\lambda$ reaches the corresponding throughput capacity ($\mu_1$, $\mu_2$ and $\mu_3$), and then per node throughput remains a constant and does not increase anymore when packet generation rate $\lambda$ goes beyond the corresponding theoretical throughput capacity. Thus, our theoretical capacity model is also efficient in depicting the per node throughput capacity behavior of the considered MANET.

\subsection{Performance Analysis}

\begin{figure}[!t]
\begin{center}
\scalebox{0.9}{\includegraphics{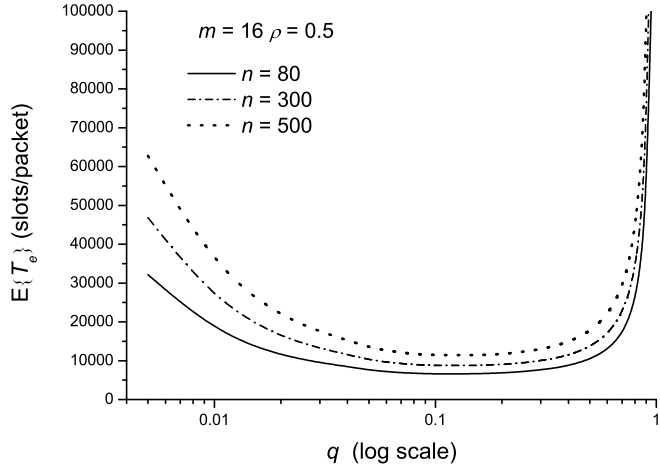}}
\end{center}
\caption{Expected packet end-to-end delay $T_e$ VS. 2HR-B parameter $q$.}
\label{fig:Teqm16graph}
\end{figure}

\begin{figure}[!t]
\begin{center}
\scalebox{0.9}{\includegraphics{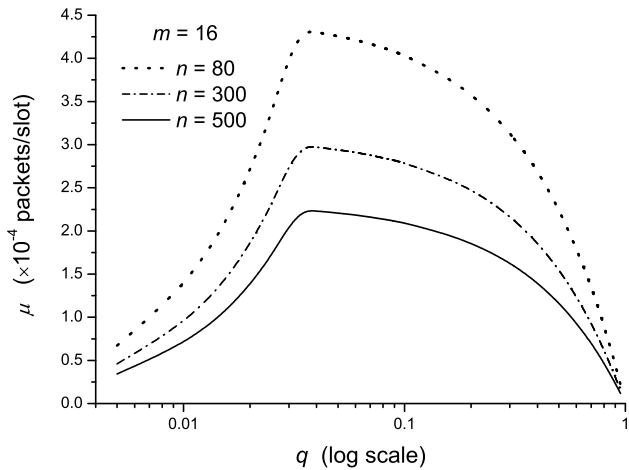}}
\end{center}
\caption{Per node throughput capacity $\mu$ VS. 2HR-B parameter $q$.}
\label{fig:muqm16graph}
\end{figure}

With the help of the QBD-based theoretical models, we explore how parameter $q$ of 2HR-B routing will affect the packet end-to-end delay $T_e$ and per node throughput capacity $\mu$ under given $m$, $n$ and $\rho$. The corresponding numerical results are summarized in Figs.~\ref{fig:Teqm16graph} and \ref{fig:muqm16graph}.

We first examine the impact of $q$ on $T_e$. For network settings of $m=16$, $\rho = 0.5$ and $n = \{80, 300, 500\}$, Fig.~\ref{fig:Teqm16graph} shows that for a given network its delay $T_e$ always first decreases and then increases as $q$ increases. This phenomenon can be explained as follows. The end-to-end delay experienced by a packet consists of the time it spends in the source-queue and the time it spends in the network-queue. An increase  in $q$ has two-fold effects on $T_e$: on one hand, it decreases the time a packet spends in the source-queue, because source $S$ has more chance to do packet-broadcast for packets in its source-queue, which makes the queue to be served more quickly; on the other hand, it increases the time a packet spends in the network-queue, because each relay has less chance to do packet-delivery to deliver a packet to destination $D$, which makes the network-queue to be served more slowly. Thus, $T_e$ decreases as $q$ increases when the first effect dominates the second one, while $T_e$ increases as $q$ increases when the second effect dominates the first one. 

We next explore how parameter $q$ affects $\mu$. For network settings of $m=16$ and $n = \{80, 300, 500\}$, Fig.~\ref{fig:muqm16graph} shows that for a given network its capacity $\mu$ always first increases and then decreases as $q$ increases. Notice that $\mu$ is determined by the minimum of service rates of the source-queue and network-queue. When $q$ is small, $\mu$ is determined by the service rate of source-queue, which increases as $q$ increases. When $q$ is large, $\mu$ is determined by the service rate of network-queue, which decreases as $q$ increases. Another observation from Fig.~\ref{fig:muqm16graph} is that the capacity $\mu$ of $n=80$ is the largest among different $n$ there. This is because that for a MANET with fixed $m=16$, a larger number of nodes there will cause a more intensive wireless channel contention, which decrease opportunities of packet transmission and thus the per node throughput capacity.

\section{Conclusion} \label{section:conclusion} 

The main finding of this paper is that the Quasi-Birth-and-Death (QBD) process  can be a promising theory  to tackle the challenging issue of analytical end-to-end delay modeling in MANETs. We demonstrated through a two-hop relay MANET that QBD theory can help us: 1) to develop a novel theoretical framework to capture the complicated network state transitions in the highly dynamic MANET, 2) to analytically model the expected end-to-end delay and also the per node throughput capacity of the network, and 3) to enable many important network dynamics like node mobility, wireless channel contention, interference and traffic contention to be jointly considered in the delay modeling process. It is expected that this work will shed light on  end-to-end delay modeling in general MANETs also.

\appendix 

\section{Proof of Lemma~2} 
\textbf{Calculation of $p_b$:} The event corresponding to $p_b$ happens iff the following sub-events happen simultaneously:

1) $S$ moves into an active cell; $j$ out of the remaining $n-1$ nodes move into the same cell with $S$, $0 \leq j \leq n-1$; other nodes move  into cells other than that active cell;

2) $S$ becomes transmitter after fair wireless channel contention;

3) $S$ selects to do packet-broadcast after traffic contention (i.e. conducting packet-broadcast or packet-delivery).

Notice that in a time slot, every node moves according to the i.i.d. mobility model. Thus, we have 
\begin{align}
p_b	&=\frac{1}{\alpha^2}\sum_{j=0}^{n-1}\binom{n\!-\!1}{j}\bigg(\frac{1}{m^2}\bigg)^j\bigg(\frac{m^2\!-\!1}{m^2}\bigg)^{n\!-1\!-j}\frac{1}{j\!+\!1}q \\
		&=\frac{qm^2}{\alpha^2n}\bigg\{ 1-\bigg(\frac{m^2-1}{m^2}\bigg)^n\bigg\}  \label{exp:pb-proof}
\end{align}

\textbf{Calculation of $p_c(j)$:} According to the definition of $p_c(j)$, it is a conditional probability determined as
\begin{align}
p_c(j)	&= \frac{p_b(j)
								}
								{p_b
								} \label{exp:pc(j)-proof}
\end{align}
where $p_b(j)$ is the probability that $j$ copies of a packet exist in the network after $S$ becomes transmitter and selects to do packet-broadcast for that packet, $1 \leq j \leq n-1$. The event corresponding to $p_b(j)$ happens iff the following sub-events happen:

1) $S$ moves into in an active cell; $j-1$ out of the remaining $n-2$ nodes other than $S$ and $D$ move into the coverage cells of $S$, among which $k$ nodes are in the same cell with $S$ and the remaining $j-1-k$ nodes are in other coverage cells of $S$, $0 \leq k \leq j-1$; other nodes move into cells other than the coverage cells of $S$;

2) $S$ becomes transmitter after fair channel contention;

3) $S$ selects to do packet-broadcast.

Notice that $D$ could move either into the same cell with $S$ or any cell other than that active cell, we have 
\begin{align}
p_b(j)	&= \!\frac{1}{\alpha^2} \bigg\{	\frac{1}{m^2}\binom{n\!-\!2}{j-1}\sum_{k=0}^{j-1}\!\binom{j\!-\!1}{k}\!\bigg(\frac{1}{m^2}\bigg)^k\!\bigg(\frac{8}{m^2}\bigg)^{j-1-k} \nonumber \\
				& \quad\quad\quad \cdot \bigg(\frac{m^2\!-\!9}{m^2}\bigg)^{n-1-j}\frac{1}{k+2}q \nonumber \\
				&	\quad\quad\quad + \frac{m^2\!-\!1}{m^2}\binom{n\!-\!2}{j-1}\sum_{k=0}^{j-1}\binom{j\!-\!1}{k}\bigg(\frac{1}{m^2}\bigg)^k \nonumber \\
				& \quad\quad\quad \cdot \bigg(\frac{8}{m^2}\bigg)^{j-1-k}\bigg(\frac{m^2\!-\!9}{m^2}\bigg)^{n-1-j}\frac{1}{k+1}q \bigg\} \\
				&=\frac{q}{\alpha^2}\binom{n-2}{j-1}\bigg(\frac{m^2-9}{m^2}\bigg)^{n-1-j} \nonumber \\
				& \quad \cdot \bigg\{\frac{m^2-9}{m^{2j}}f(j)+\frac{1}{m^{2j}}f(j+1)\bigg\}  \label{exp:pb(j)-proof}
\end{align}
where
\begin{align}
f(x)	&= \frac{9^x-8^x}{x}
\end{align}
After substituting (\ref{exp:pb(j)-proof}) and (\ref{exp:pb-proof}) into (\ref{exp:pc(j)-proof}) and conducting some basic algebraic calculations, we have
\begin{align}
p_c(j)	&=	\!\frac{n\binom{n-2}{j-1}(m^2\!-\!9)^{n-1-j}}{m^{2n}-(m^2-1)^n}\Big\{(m^2-9)f(j)+f(j\!+\!1)\Big\}	
\end{align}

\textbf{Calculation of $p_r(j)$:} Notice that in a time slot, $D$ could only receive the packet it is currently requesting from one of the $j$ nodes carrying copies of that packet. Thus, after similar arguments to the calculation of $p_b$, we have
\begin{align}
p_{r}(j)		&=	\!\frac{j(1\!-\!q)m^2}{\alpha^2n(n\!-\!1)} \bigg\{	1\!-\!\bigg(\frac{m^2\!-\!1}{m^2}\bigg)^n\!-\!\frac{n}{m^2}
									\bigg(\frac{m^2\!-\!9}{m^2}\bigg)^{n\!-\!1}\!\bigg\}	
\end{align}

To calculate the remaining probabilities, we need to construct the arrival process of network-queue. From Lemma~1, we know that the arrival process of network-queue is a Bernoulli process with probability $\lambda$. The arrival process of network-queue can be constructed as follows: once source node $S$ becomes transmitter and selects to do packet-broadcast (with probability $p_b$), $S$ successfully conducts packet-broadcast for one packet with probability $\lambda'$, 
\begin{align}
\lambda'	&=	\frac{\lambda}{p_b}
\end{align}
Thus, after $S$ becomes transmitter and selects to do packet-broadcast, $S$ will successfully distribute out one packet with probability $\lambda'$.

\textbf{Calculation of $p_{0}(j)$:} The event corresponding to $p_{0}(j)$ happens iff the following sub-events happen:

1) $S$ moves into an active cell; $D$ moves into any cell other than that active cell; $j-1$ out of the remaining $n-2$ nodes move into the coverage cells of $S$, among which $k$ nodes are in the same cell with $S$ and the remaining $j-1-k$ nodes are in other coverage cells of $S$, $0 \leq k \leq j-1$; other nodes move into cells other than the coverage cells of $S$;

2) $S$ becomes transmitter after fair contention;

3) $S$ selects to do packet-broadcast and $S$ distributes out a packet.

Then, we have 
\begin{align}
p_{0}(j)	&=	\!\frac{1}{\alpha^2}\frac{m^2\!-\!9}{m^2}\binom{n\!-\!2}{j\!-\!1}\!\sum_{k=0}^{j-1}\!\binom{j\!-\!1}{k}\!\bigg(\frac{1}{m^2}\!\bigg)^k\!\bigg(\frac{8}{m^2}\bigg)^{j-1-k} \nonumber \\
							& \quad \cdot \bigg(\frac{m^2-9}{m^2}\bigg)^{n-1-j}\frac{1}{k+1}q\lambda' \\
							&=	\frac{\lambda \cdot q \cdot \binom{n-2}{j-1}(m^2-9)^{n-j}}{\alpha^2 m^{2n-2}p_{b}}f(j)  \label{exp:p0j-proof}
\end{align}

\textbf{Calculation of $p_{0}(0)$:} From the definition of $p_{0}(0)$, we know that 
\begin{align}
p_{0}(0)&= 1 - \sum_{j=1}^{n-1}p_{0}(j)	 \label{exp:p00-p0jrelation}
\end{align}
After substituting (\ref{exp:p0j-proof}) into (\ref{exp:p00-p0jrelation}), we have 
\begin{align}
p_{0}(0)	&=	1- \frac{\lambda \cdot q \cdot (m^2-9)}{\alpha^2(n-1)p_{b}}\bigg\{	1-\bigg(\frac{m^2-1}{m^2}\bigg)^{n-1}\bigg\}
\end{align}

\textbf{Calculation of $p_{b}^{+}(j)$:} The event corresponding to $p_{b}^{+}(j)$ is composed of $j-1$ exclusive sub-events, each of which is that: in a time slot $S$ becomes transmitter, selects to do packet-broadcast for one packet; at the same time $D$ receives the packet it is requesting from a specific relay node (say $R$) carrying a copy of that packet. If we denote by $p_{b}^{+}$ the probability that one such sub-event occurs in a time slot, then 
\begin{align}
p_{b}^{+}(j)	&=	(j-1)p_{b}^{+}	\label{exp:pb+j-proof}
\end{align}
The event corresponding to $p_{b}^{+}$ happens iff the following sub-events happen:

1) $S$ moves into an active cell; $R$ moves into another active cell; $D$ moves into either the same active cell with $R$ or other coverage cells of $R$; $k$ out of the remaining $n-3$ nodes move into the coverage cells of $R$, among which $i \geq 0$ nodes are in the same cell with $R$; $t \geq 0$ of the remaining $n-3-k$ nodes are in the same active cell with $S$; other nodes move into cells other than the active cell of $S$ and the coverage cells of $R$;

2) $S$ and $R$ both become transmitters after fair contention in their respective active cells;

3) $S$ selects to do packet-broadcast and $S$ distributes out a packet; $R$ selects to do packet-delivery and $D$ is selected as its receiver.

Then, we have
\begin{align}
p_{b}^{+}	&=	\!\frac{1}{\alpha^2}\frac{m^2\!-\!\alpha^2}{m^2\alpha^2}\sum_{k=0}^{n-3}\!\binom{n\!-\!3}{k}\!\Bigg\{\!\sum_{i=0}^{k}\binom{k}{i}\!\bigg(\!\frac{1}{m^2}\!\bigg)^i\!\bigg(\!\frac{8}{m^2}\!\bigg)^{k-i} \nonumber \\
				&		\cdot	\sum_{t=0}^{n-3-k}\binom{n\!-\!3\!-\!k}{t}\bigg(\frac{1}{m^2}\bigg)^t\bigg(\frac{m^2-10}{m^2}\bigg)^{n-3-k-t} \nonumber \\
				&		\cdot \! \bigg(\frac{1}{m^2}\frac{1}{i+2}\frac{1}{k+1}+\frac{8}{m^2}\frac{1}{i+1}\frac{1}{k+1}\bigg)\!(1\!-\!q)\frac{1}{t\!+\!1}q\lambda'\!\Bigg\} \\
				&=\frac{\lambda(q-q^2)(m^4-m^2\alpha^2)}{\alpha^4 n(n-1)(n-2) p_{b}}	\nonumber \\
							&		\quad \cdot \bigg\{1-2\bigg(\frac{m^2-1}{m^2}\bigg)^n + \bigg(\frac{m^2-2}{m^2}\bigg)^n	\nonumber \\
							&		\quad		-\frac{n}{m^2}\bigg(\frac{m^2-9}{m^2}\bigg)^{n-1}\!+\!\frac{n}{m^2}\bigg(\frac{m^2\!-\!10}{m^2}\bigg)^{n\!-\!1}\bigg\}	\label{exp:pb+}
\end{align}
From (\ref{exp:pb+}) and (\ref{exp:pb+j-proof}), (\ref{exp:pb+(j)}) follows.

\textbf{Calculation of $p_{b}^{-}(j)$:} From the definitions of $p_{b}^{-}(j)$ and $p_{b}^{+}(j)$, we know that $p_{b}^{-}(j) + p_{b}^{+}(j)$ is the probability that in a time slot $S$ becomes transmitter, selects to do packet-broadcast and also successfully conducts packet-broadcast for one packet. This probability is just $\lambda$ according to the arrival process of network-queue. Thus, $p_{b}^{-}(j)$ can be calculated as 
\begin{align}
p_{b}^{-}(j)	&= \lambda - p_{b}^{+}(j)
\end{align}

\textbf{Calculation of $p_{f}^{+}(j)$:} By the definition of $p_r(j)$, it is easy to see that $p_{f}^{+}(j)$ can be calculated as 
\begin{align}
p_{f}^{+}(j)	&=	p_r(j) - p_{b}^{+}(j)
\end{align}

\textbf{Calculation of $p_{f}^{-}(j)$:} From the definitions of $p_{b}^{+}(j)$, $p_{b}^{-}(j)$, $p_{f}^{+}(j)$ and $p_{f}^{-}(j)$, we know that
\begin{align}
p_{b}^{+}(j) + p_{b}^{-}(j) + p_{f}^{+}(j) + p_{f}^{-}(j)	&=	1
\end{align}
Thus, (\ref{exp:pf-(j)}) follows.

\bibliography {reference}
\end{document}